\documentclass[aip, reprint]{revtex4-2}

\usepackage{amsmath}
\usepackage{amsfonts}
\usepackage{amssymb}

\usepackage{graphicx}
\usepackage{dcolumn}
\usepackage{bm}

\usepackage[utf8]{inputenc}
\usepackage[T1]{fontenc}
\usepackage{mathptmx}

\usepackage{subfigure}
\usepackage{upgreek}
\usepackage[usenames]{color}

\newcommand{\Fi}{\mathcal F}

\begin{document}
\title{Mapping standing-wave cavity modes with a commercial scanning near-field microscope tip}

\author{Francesco Ferri}
\altaffiliation[Present address: ]{Institute for Quantum Electronics, ETH Z\"{u}rich, 8093 Z\"{u}rich, Switzerland}
 \email{fferri@phys.ethz.ch.}
\author{S{\'e}bastien Garcia}
\altaffiliation[Present address: ]{Coll\`{e}ge de France, 11 Place Marcelin Berthelot, 75321 Paris,
France.}
\author{Mohamed Baghdad}
\author{Jakob Reichel}
\author{Romain Long}
 \email{long@lkb.ens.fr.}
\affiliation{Laboratoire Kastler Brossel, ENS-Universit\'{e} PSL, CNRS, Sorbonne Universit\'{e}, Coll\`{e}ge de France, 24 rue Lhomond, 75005 Paris, France}

\date{\today}

\begin{abstract}
We describe a method to map the standing-wave pattern inside a Fabry-Perot optical cavity with sub-wavelength resolution by perturbing it with a commercially available scanning near-field optical microscope (SNOM) tip. The method is applied to a fiber Fabry-Perot microcavity. We demonstrate its use to determine the relative position of the antinodes at two different wavelengths. In addition, we use the SNOM tip as a point-like source allowing precise positioning of a microscope objective with respect to the cavity mode.
\end{abstract}

\maketitle

\section{Introduction}

Measuring the standing-wave intensity distribution inside a free-space optical cavity is a challenging task that has practical importance for enhancing the field-emitter coupling in cavity quantum electrodynamics (CQED) and its applications in quantum technologies, such as single-photon sources \cite{Eisaman2011} and quantum interfaces.\cite{Heshami2016} 
Knowledge of the intensity distribution is important for optimum positioning of a quantum emitter inside the cavity, and for determining the emitter-field coupling that can be achieved. 
Factors such as deviation of the mirror shape from an ideal sphere, or wavelength-dependent penetration depth of the field into the multilayer dielectric mirror may lead to significant deviations of the intracavity field from a simple calculated distribution.\cite{Benedikter2015} In addition, in many CQED and quantum interface experiments with neutral atoms, \cite{Hosten2016, Zhiqiang2017, Davis2019, Garcia2018} two standing waves of different wavelengths are present in the cavity, one forming an optical lattice to trap the atoms and the other providing resonant or slightly detuned coupling. Here, the relative positioning  of the two fields determines the coupling strength at a given lattice site. 
For recent atomic CQED experiment, \cite{ Welte2017, Leonard2017, Gallego2018, Davis2019} the cavity mode also has to be precisely positioned with respect to other external optical elements such as high numerical aperture (NA) objectives.  In this context, the ability to not only map the standing wave but also reveal its absolute position is especially relevant. 

Since the pioneering experiment of Wiener,\cite{Wiener1890} probing standing-wave patterns has been achieved by using specific photodetectors thinner than the optical wavelength,\cite{Sasaki1999} which are now used in standing-wave sensors and spectrometers.\cite{Stiebig2006} However, this method cannot be implemented in a high-finesse cavity due to the large losses introduced by the detector.
In whispering-gallery-mode resonators and photonic crystal cavities, the cavity field can be measured by probing the evanescent field with scanning near-field optical microscopes (SNOMs).\cite{Knight1996, Gotzinger2001, Koenderink2005, Rotenberg2014} But this approach is not applicable to open, three-dimensional cavities such as the Fabry-Perot resonators (unless adding a prism inside a low-finesse cavity\cite{Courjon1994}).
In a landmark experiment in 2001,\cite{Guthohrlein2001} a trapped, laser-cooled ion was employed to map out the standing-wave pattern of a high-finesse Fabry-Perot cavity in vacuum. 
More recently in the context of quantum optomechanics,  custom-made high-frequency vibrating nanorod oscillators gave access to spatially resolved field information.\cite{Favero2009, Gloppe2014, Fogliano2019}
Another possibility for detecting intra-cavity optical standing waves in an optomechanics setting is to use nanomembranes, whose effect is to induce a position-dependent dispersive shift and  losses,\cite{Thompson2008, Vochezer2018} with the limitation to provide only a one-dimensional mapping of the optical mode and to require large optical access.

\begin{figure}
\centering 
\includegraphics[width=1\columnwidth]{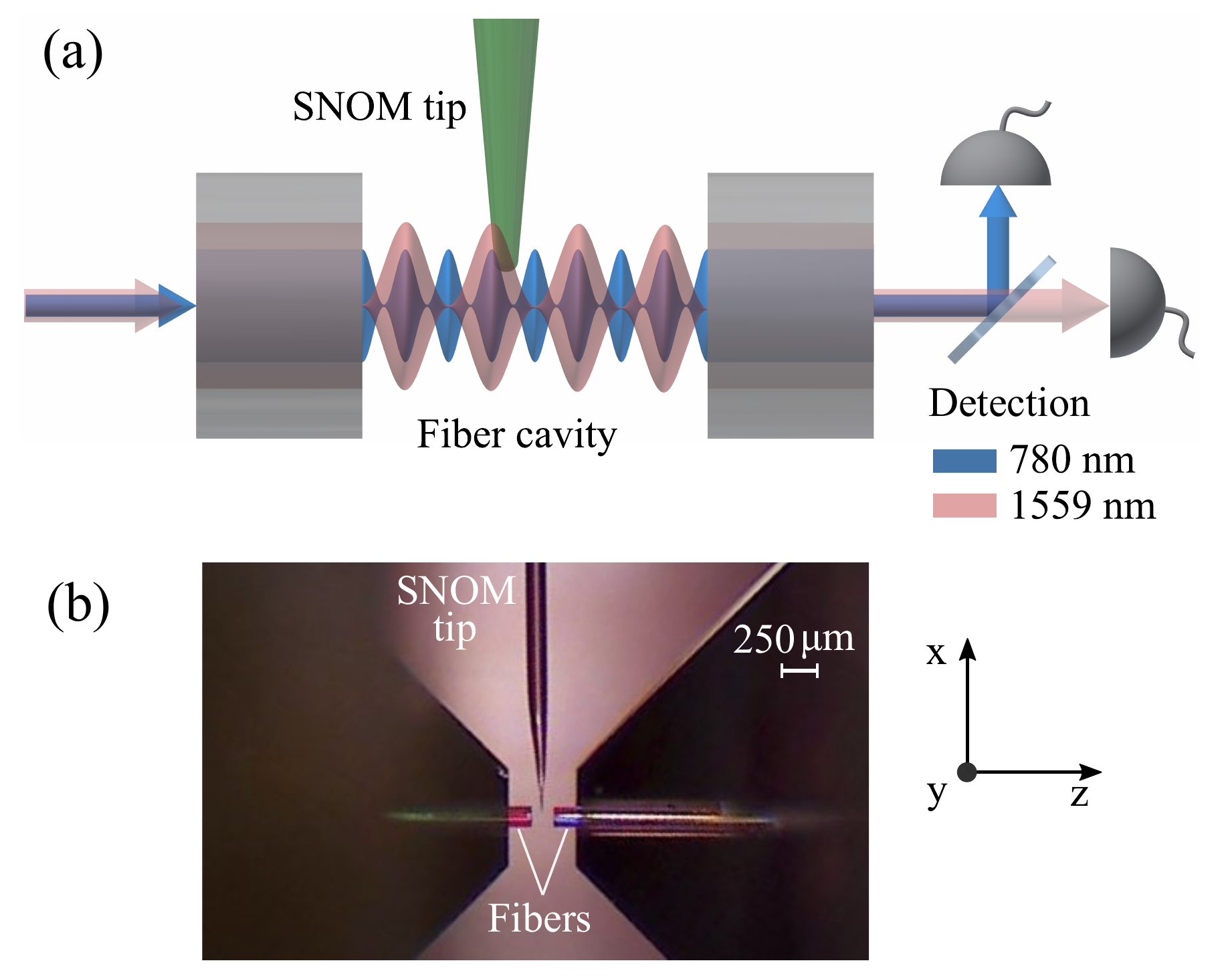}
\caption{(a) Schematic of the experiment to map the modes of a fiber Fabry-Perot by using a SNOM tip (shown in green). The tip induces spatially dependent losses on two TEM$_{00}$ modes at $780\,$nm (blue) and $1559\,$nm (red), which are detected by measuring the transmission spectra of the cavity on separate photodetectors. Dimensions are not to scale. (b) Image via a low-resolution microscope showing the fiber cavity (horizontally oriented) and the SNOM tip (vertically oriented).}
\label{fig:Figure1}
\end{figure}

Here, we describe a standing wave mapping method for Fabry-Perot or other open resonators based on the perturbation of the cavity mode by a commercially available sub-wavelength sized tip. The tip is constituted by a pulled optical fiber and intended for scanning near-field optical microscopy (SNOM). We use this method to measure the relative positioning of two standing-wave fields one octave apart inside a fiber Fabry-Perot (FFP) microcavity.\cite{Hunger2010, Garcia2018}
More specifically, our microcavity setup aims at enabling maximum and uniform coupling between cold rubidium atoms and a cavity mode at a wavelength $\lambda_1$ close to the $D_2$ line of $^{87}$Rb at 780\,nm. As the atoms are optically trapped  at the antinodes of an intra-cavity one-dimensional lattice with a wavelength $\lambda_2\approx 2\lambda_1$, an essential requirement is that the antinodes for $\lambda_2$ overlap with the antinodes of $\lambda_1$. In this context, the SNOM tip provides a simple tool to verify the correct overlap of the antinodes of the two commensurable standing waves. 

In addition, we used a coated SNOM tip to align a high-NA objective onto the cavity mode. The tip can indeed be used as a nearly perfect point source, by coupling light into the optical fiber of the tip. At the same time, monitoring the tip-induced cavity losses allows to pinpoint the absolute position of the cavity mode, onto which the objective can be aligned. Such use of the SNOM tip as a nanoscopic lighthouse finds important applications in experimental designs where the cavity mode has to be precisely positioned with respect to other optical elements.

\section{Measurement principle}
\label{text:Tip}

Our method for visualizing intra-cavity standing waves is based on the use of a pointed probe object with sub-wavelength tip size. When the probe is inserted in the cavity, it introduces additional optical losses that depend on the local overlap between the tip and the cavity mode (see Fig.~\ref{fig:Figure1}, not to scale). This perturbation can be detected as variations of the amplitude and of the spectral width of the cavity resonance. Thus, moving the probe along the cavity axis allows reconstructing the spatial distribution of the optical standing waves. 

More quantitatively, the round-trip losses introduced by the probe $\mathcal{L}\left(x,y,z\right)$ depend on the tip position given by $(x,y,z)$, where $z$ is the coordinate along the cavity axis (see Fig.~\ref{fig:Figure1}). 
When the tip is displaced along $z$ at constant $x$ and $y$, the losses will change periodically between a minimum and a maximum value, corresponding to the positions of respectively the nodes and antinodes of the intra-cavity standing waves.
These losses add to the intrinsic cavity losses, such as transmission, scattering and absorption of the mirror coatings.
The on-resonance intensity cavity transmission $T$ and the finesse $\mathcal{F}$ are then modified as follows:
\begin{equation}\label{tip_transmission}
\frac{T(x,y,z)}{T_{0}}=\left(1+\frac{\mathcal{F}_{0}}{2\pi}\mathcal{L}_{}\left(x,y,z\right)\right)^{-2}
\end{equation}
\begin{equation}\label{tip_finesse}
\frac{\mathcal{F}(x,y,z)}{\mathcal{F}_{0}}=\left(1+\frac{\mathcal{F}_{0}}{2\pi}\mathcal{L}_{}\left(x,y,z\right)\right)^{-1}
\end{equation}
where $T_{0}$ and $\mathcal{F}_{0}$ are respectively the unperturbed values of the transmission and of the finesse. 

From the measurement of these quantities, we can directly reconstruct the intra-cavity standing wave pattern given that the eigenmodes of the resonator are not significantly modified by the tip. In our measurement, the dispersive shift due to the tip is  indeed about $10^{-3}$ times smaller than the free spectral range.  Thus, we can consider that the losses induced by the probe directly reflect the spatial distribution of the unperturbed optical fields.

\section{Experimental Setup}

\subsection{Fiber Fabry-Perot cavity}
\label{sec:FFPC}

The combination of optical cavities with ultracold atoms is one of the promising avenue for the generation of multiparticle entangled states. \cite{Barontini2015, McConnell2015} To reach the strong coupling regime of CQED and to add individual atoms detection capabilities, we have built a high-finesse fiber Fabry-Perot microcavity operating with high finesse at the two wavelengths of $\lambda_1=780\,$nm and $\lambda_2=1559\,$nm. \cite{Garcia2018}
The cavity length is about $L = 135\,\upmu$m and the fiber mirrors are close to be perfectly spherical  with a radius of curvature of $R= 315\,\upmu$m, see Ref.~\onlinecite{Garcia2018}. The calculated waists at the two wavelengths are $w_{1}=5.6\,\upmu$m and $w_{2}=8.1\,\upmu$m, and the unperturbed cavity finesses are $\Fi_{1}=(4.5\pm0.2)\times10^4$ and  $\Fi_{2}=(7.9\pm0.3)\times10^4$. 
The cavity has been designed to realize maximum spatial overlap (superposition of the antinodes) between the modes at $\lambda_1$ and $\lambda_2$ by choosing an appropriate phase shift between the modes at their reflection on the dielectric mirrors. The opposite (in our case, unwanted) configuration of minimal overlap
(coincident nodes) occurs for the same
optical frequencies but with a different relative phase at reflection, and therefore it cannot be distinguished spectroscopically from the desired one. The mapping method presented here resolves the two different configurations by providing a direct detection of the spatial distribution of the cavity modes.

To keep the setup simple, we avoid a cavity locking servo by combining reasonable passive stability with a sweeping scheme. The cavity length is constantly modulated by about $\pm 100\,$nm (which is more than the amplitude of mechanical vibrations) at $20\,$Hz by means of a piezoelectric actuator. After splitting the path of the two wavelengths with a dichroic mirror, we detect the cavity transmission at $\lambda_1$ and $\lambda_2$ on two independent photodiodes.

\subsection{SNOM tip}
To probe the standing wave, we use a commercially available tapered optical fiber (Lovalite, Besan\c con - FR) with an apex radius of $(50 \pm 20)\,$nm , which is an intermediate product in the fabrication of tips for scanning optical near-field microscopy. This probe is inexpensive and its manipulation does not require much more care than a standard optical fiber without coating. 

The tip is placed on a three-axis translation stage (Thorlabs MBT616D/M) and positioned inside the cavity. Two piezo actuators allow moving the tip along the direction of the cavity axis $z$ and along the direction of the probe tip axis $x$ (Fig.~\ref{fig:Figure1}). We adjust the $y$ position of the tip to place it at the center of the cavity modes. This can be done by finding the minimum transmission while moving the tip along $y$.  
To make sure that $z$ axis displacement of the tip is parallel to the axis of the cavity mode, we use the cavity transmission as an indicator: if the motion is not parallel, the $\lambda/2$-modulated cavity transmission has a nonzero global slope, due to the fact that the insertion depth of the tip in the mode changes during the motion. We manually adjust the angle of the tip relative to the cavity until a minimum slope is reached over the whole range of the $z$-axis piezo actuator of the probe ($\pm10\,\upmu$m around the center of the cavity). Following this procedure, we estimate the residual alignment error to be less than $0.2^{\circ}$.
 
\subsection{Data acquisition}
Once the tip is aligned inside the cavity, we slowly sweep its $z$ position by $\pm0.7\,\upmu$m with the piezo actuator. Sweep duration is $100-250\,$s (i.e., much longer than the cavity modulation period). At the same time, we keep modulating the cavity length as before and we record the transmission spectra at $\lambda_{1}$ and $\lambda_{2}$, as well as the voltage on the tip piezo actuator (proportional to the displacement along $z$) with an oscilloscope. In this way, we obtain a series of cavity spectra corresponding to different $z$ positions of the tip. From these spectra, the values for the perturbed on-resonance transmission and finesse are extracted. Both the transmission and the finesse allow to extract the tip-induced losses $\mathcal{L}$ (Eq.~\eqref{tip_transmission}, Eq.~\eqref{tip_finesse}). However, the measurement of the finesse via the resonance linewidth is more affected by 
fast mechanical vibrations during the scan that slightly distort the line shape. In the following, we will present only data based on the measurement of the on-resonance transmission, which provides better signal-to-noise ratio for $\mathcal{L}$.

The tip position along $z$ is adjusted through the voltage applied to the piezo actuator of the tip. To calibrate this displacement, we use as a reference the periodic modulation of the cavity transmission due to the standing wave at $\lambda_1=780.2\,$nm (wavelength of the D$_2$ line of $^{87}$Rb, to which the laser is set). From the comparison with a 5-periods standing wave signal at $1559\,$nm we also estimate a non linear error of the piezo scan of about $10\%$. All acquisitions are performed while the tip moves in the same direction, in order to avoid considering the hysteresis of the piezo actuator. The reproducibility of the tip scans along $z$ is mostly limited by vibrations and thermal drifts. By restricting the time for data taking to less than one hour, we observe that the position of the tip relative to the maxima of the optical modes along $z$ can be inferred within a precision of $\pm30\,$nm.

\section{Mapping standing waves}

\subsection{Measurement of the overlap between two standing waves}

\begin{figure}
\centering
\includegraphics[width=\columnwidth]{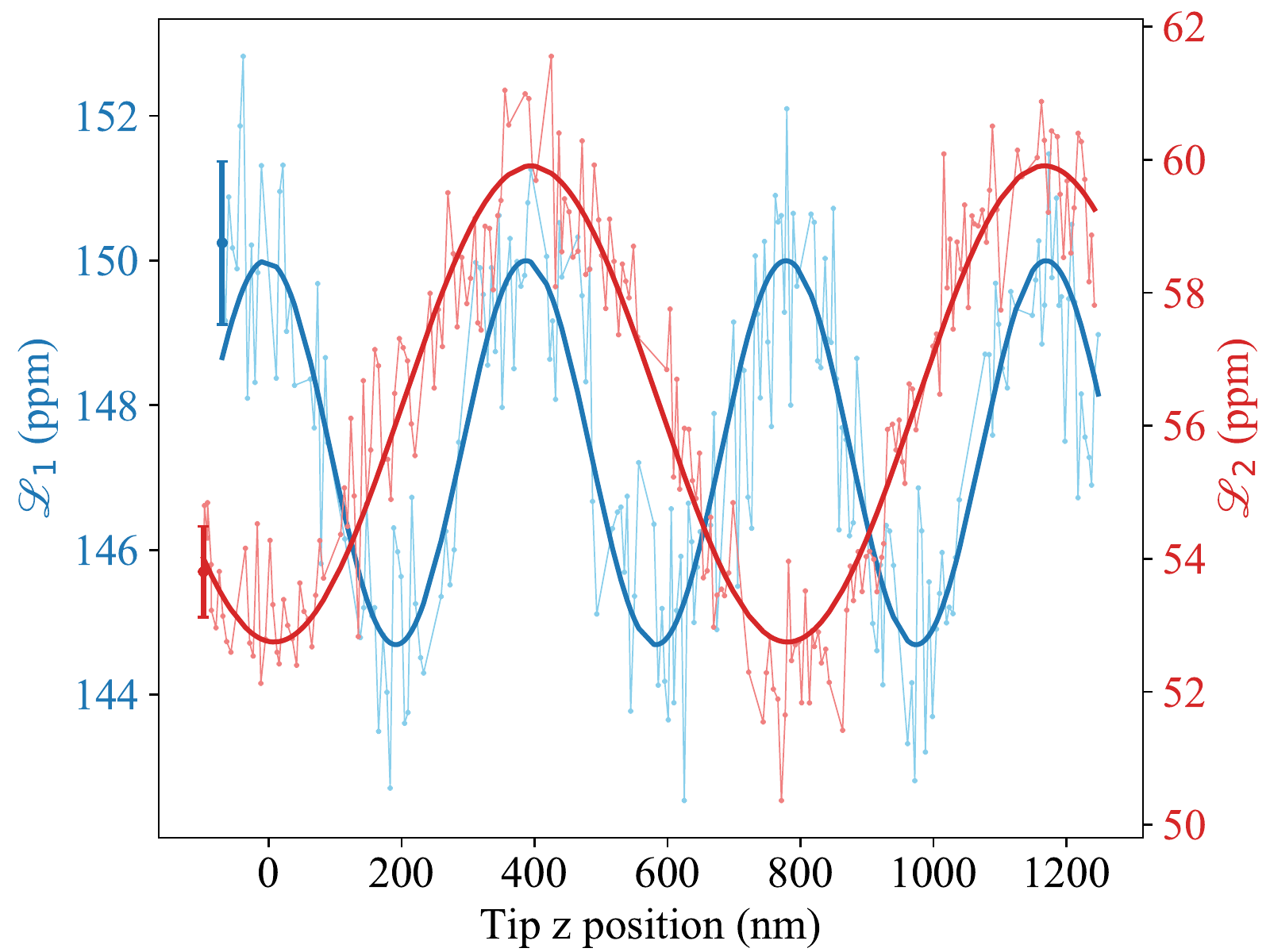}
\caption{Tip-induced optical losses at $\lambda_{1}=780\,$nm (blue) and $\lambda_{2}=1559\,$nm (red) as a function of the tip position along the cavity axis, in which the imprint of the spatial structure of the standing waves is clearly visible. Experimental data are shown in light colors. We associate to each data point an error bar (presented only once in the plot) which we estimate as the fluctuations of the signal (standard error of the mean over 6 points) when the tip is kept fixed at the same depth. Thick lines are best-fit curves obtained from Eq~\eqref{eq:fit_losses}.}
\label{fig:Figure3}
\end{figure}

In Fig.~\ref{fig:Figure3}, we show the result of two sweeps of the tip along the cavity axis, one for each wavelength $\lambda_{1}$ and $\lambda_{2}$. Each point corresponds to a value of the cavity losses extracted from the on-resonance cavity transmission using Eq~\eqref{tip_transmission}. The two sweeps are performed at different insertion depths ($x$ positions) of the tip, which are chosen to optimize the signal-to-noise ratio for the standing wave detection (see \ref{sec:snratio}).

The solid lines in Fig.~\ref{fig:Figure3} are best-fit curves obtained by assuming the following simple model for the optical losses as a function of $z$:
\begin{equation}\label{eq:fit_losses}
\mathcal{L}_{i}(z)=\overline{\mathcal{L}}_i+\Delta \mathcal{L}_i\cos\left[\frac{4\pi}{\lambda_i}(z-z_{0,i})\right],
\end{equation}
with four free parameters (mean losses $\overline{\mathcal{L}}_i$, loss variations with amplitude $\Delta \mathcal{L}_i$, wavelength $\lambda_i$ and spatial offset $z_{0,i}$) for each wavelength $i=1,2$. 
The fit function used here to model the losses variation allows us to fit our data and to extract information of interest about the standing waves and their overlap.

From the best-fit parameters, we obtain $\lambda_{2}/\lambda_{1} = 2.0 \pm 0.1$, where the uncertainty is dominated by the imperfect reproducibility of the tip scans along $z$. The fit positions of the loss maxima at $\lambda_{2}$ and $\lambda_{1}$ coincide to better than $1\,$nm (which is partly fortuitous given the absolute position uncertainties of $\pm 30\,$nm). We can then conclude that the antinodes of the standing waves are coincident, as intended by cavity design (see \ref{sec:FFPC}). For completeness, we also get the following values for the losses due to the tip: $\overline{\mathcal{L}}_{1}=(147\pm 6)\,$ppm, $\Delta\mathcal{L}_{1}=(2.65\pm0.11)\,$ppm for $\lambda_{1}$; $\overline{\mathcal{L}}_{2}=(56\pm2)\,$ppm, $\Delta\mathcal{L}_{2}=(3.57\pm0.07)\,$ppm for $\lambda_{2}$.

As visible in Fig.~\ref{fig:Figure3}, we limit the measurement to only a few half-periods of the standing waves. The range of the $z$-scan is the result of a compromise between acquiring enough data for a reliable reconstruction of the periodic signal and performing the measurement in a short time to avoid drifts. The main noise source affecting the measured traces is indeed the vibrations of the cavity and of the tip. Our setup has been designed to allow high flexibility during the testing phase and the assembly of the fiber cavity; many improvements could be easily implemented to reduce mechanical vibrations, enabling better detection of standing waves. However, the fact that useful data can be obtained without such optimizations emphasizes the simplicity and robustness of the method presented here, which does not require strict conditions and can then constitute an useful tool for the realization of new optical setups.

\subsection{Optimization of the mapping}
\label{sec:snratio}

\begin{figure}[thbp]
\centering
\includegraphics[width=\columnwidth]{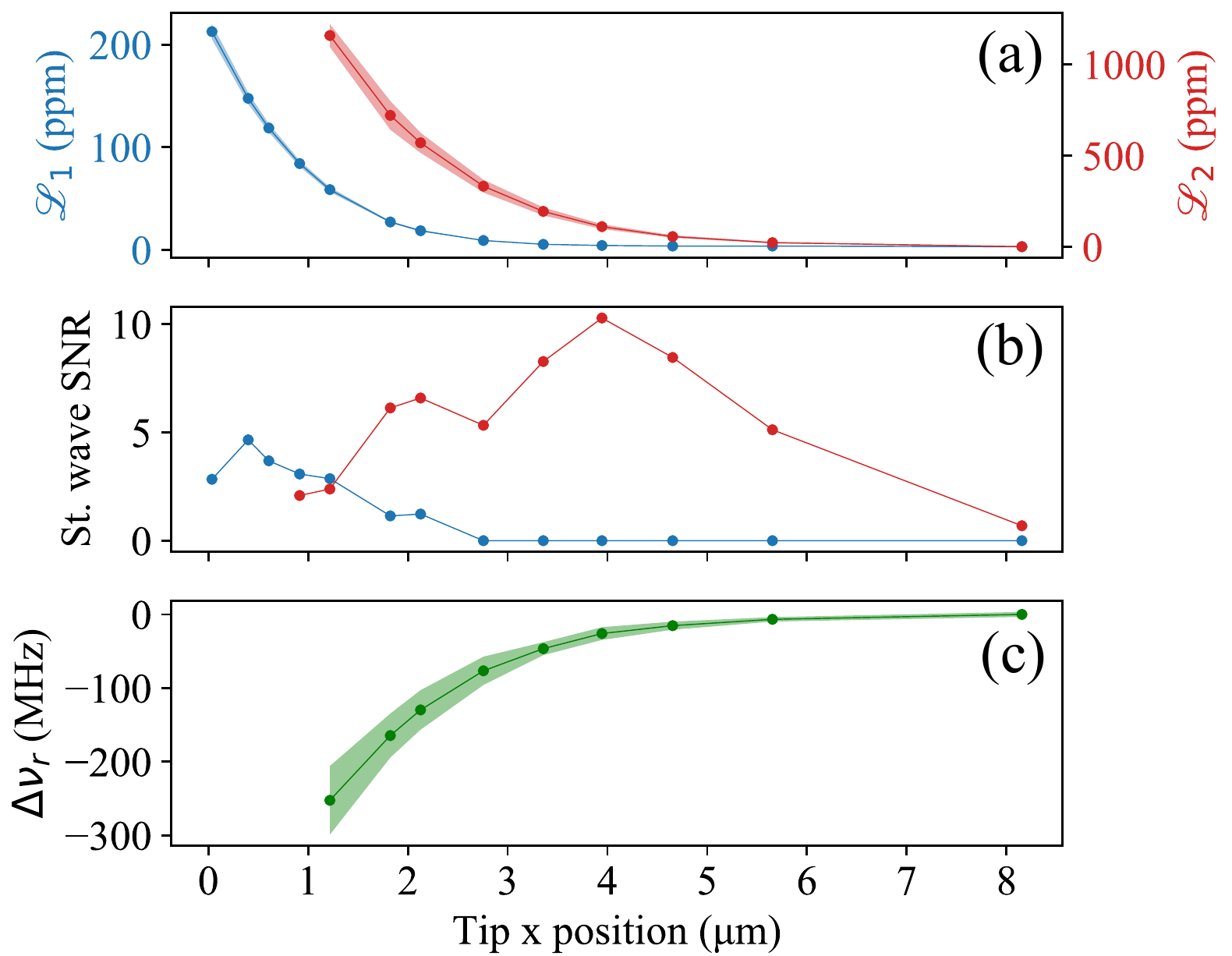}
\caption{(a) Optical losses induced by the tip for the modes at $\lambda_1=780\,$nm (blue) and $\lambda_2=1559\,$nm (red), as a function of the depth of the tip into the cavity mode ($x$ position). Higher values of $x$ correspond to more peripheral positions and the cavity center is estimated to be near $x=0$. (b) Measured signal-to-noise ratio (SNR) for the standing waves detection, as a function of the tip depth $x$ (color code is the same as in (a). The SNR is defined by the ratio between the amplitude of the $\lambda/2$ modulation, captured by the parameter $\Delta\mathcal{L}$ of the fitting function Eq.~\eqref{eq:fit_losses} and the standard deviation of the fluctuations from the fitted signal. (c) Variation of the relative dispersive shift $\Delta\nu_r$ between the two modes (see main text for details) as a function of $x$. In (a), (c) the dots are averages over a period $\lambda_{1,2}$ (detected by moving the tip along the cavity axis $z$ at constant $x$), while the shaded regions show the amplitude of the $z$ standing-wave modulation, where visible.}
\label{fig:Figure2}
\end{figure}

In our method, the choice of the immersion depth ($x$ position) of the tip into the cavity mode plays a crucial role. As we will discuss, it affects the signal-to-noise ratio of the standing wave detection. In addition, it determines the overall dispersive shift due to the tip, whose value is important for a correct interpretation of the position-dependent optical losses. We report in Fig.~\ref{fig:Figure2}(a) detailed analysis of the perturbation due to the tip as a function of its depth into the optical modes. The data are obtained by measuring the cavity spectrum while sweeping the tip along $z$, as in Fig.~\ref{fig:Figure3}, for different tip depths $x$. The origin $x=0$ is set to the estimated position of the optical axis of the cavity, and higher values of $x$ correspond to more peripheral positions of the tip. Fig.~\ref{fig:Figure2}(a) shows the induced losses $\mathcal{L}_{1}$ and $\mathcal{L}_{2}$ for the two wavelengths. Dots are the  losses averaged over one period of the standing wave, while the shaded regions represent the amplitude of the standing wave modulation for each value of the tip depth. As expected, the perturbation due to the tip is larger for increasing spatial overlap with the cavity modes. 

The variation of the losses with the $x$ position shown in Fig.~\ref{fig:Figure2}(a) has an important influence on the signal-to-noise ratio (SNR) for the detection of standing waves, reported in Fig.~\ref{fig:Figure2}(b). We observe the existence of an optimal depth, distinct for the two wavelengths, which maximizes the SNR. 
The presence of an optimum at intermediate depth can be understood from Eq.~\eqref{tip_transmission}: for $\mathcal{L}\ll\mathcal{L}_{0}$ (where $\mathcal{L}_0=2\pi/\mathcal{F}_0$ are the losses of the unperturbed cavity), the effect of the tip is too small
to be detectable above the acoustic noise of the cavity transmission.
On the other hand, for $\mathcal{L}\gg\mathcal{L}_{0}$, the perturbation drastically reduces the transmission of the cavity, making photodetectors noise the leading noise source. 

The optimal SNR is obtained at different depth of the tip for $\lambda_{1}$ and $\lambda_{2}$ (Fig.~\ref{fig:Figure2}(b)). This is due to the fact that the two corresponding cavity modes have different waist sizes and different periodicity. We also see that, as expected, the standing wave at $\lambda_{2}$, where the distance between the maxima is twice as large as for $\lambda_{1}$, can be detected with a better signal-to-noise. As a consequence, within our experimental framework, it is nearly impossible to detect the standing waves at $\lambda_{1}$ and $\lambda_{2}$ simultaneously during the same sweep of the tip. However, the alignment procedure described above allows us to compare measurements taken at different insertion depth of the tip. By taking into account the uncertainty of the alignment procedure ($0.2^{\circ}$) and temporal fluctuations of the angle between the tip and the cavity (also on the order of $0.2^{\circ}$), we estimate indeed that the
cross-talk between the $z$ and the $x$ displacement of the tip is $\delta z/\delta x < 5\,$nm/$\upmu$m. Experimentally, we have compared the standing waves for the same wavelength but measured with different tip depths within the visibility range. We have not observed deviations of the maxima positions due to initial misalignment within the uncertainty. The two standing wave traces shown in Fig.~\ref{fig:Figure3} have been acquired by setting tip depth $x$ such that the SNR is close to maximum for each wavelength, i.e. at $0.4\,\upmu$m for $780\,$nm and $4.6\,\upmu$m for $1559\,$nm.

Lastly, we consider the dispersive effect due to the tip by referring to Fig.~\ref{fig:Figure2}(c). Due to vibrations and drift, the voltage of the piezo that modulates the cavity length is not a good measure of absolute cavity length. We can detect only the length difference between the two resonances for $\lambda_1$ and $\lambda_2$. We define the relative frequency shift between the two resonances as $\nu_r=2\nu_2-\nu_1$, where $\nu_{1,2}$ are the frequencies of the cavity modes  at $\lambda_{1,2}$. For this measurement, the tip depths $x$ are restricted to values that allow to detect simultaneously the two resonances at  $\lambda_1$ and $\lambda_2$. In these conditions, only the standing wave modulation at $\lambda_2$ is resolved and the dispersive shift is mainly due to the variation of $\nu_2$. The shift $\nu_r$ is measured from the cavity spectra by using the sidebands of an electro-optical modulator on the $\lambda_1$ line as a frequency reference. In Fig.~\ref{fig:Figure2}(c) we show the variation of the relative dispersive shift $\Delta\nu_r=\nu_r-\nu_{r,0}$ as a function of the tip depth $x$ ($\nu_{r,0}$ is the unperturbed value of the relative frequency between the two modes). When sweeping the tip along the optical axis $z$ at constant $x$, a modulation of $\Delta\nu_r$ with period $\lambda_2/2$ is visible. We represent with dots are averages over a modulation period, and with a shaded region the amplitude of this modulation. Within the range of tip depth we explored, $\Delta\nu_r$ scales approximately linearly with the induced losses at $1559\,$nm $\mathcal{L}_{2}$. The induced dispersive effect is about $10^{-3}$ smaller than the free spectral range of the cavity, which makes the perturbation on the spatial distribution of the optical modes along $z$ negligible in the precision of our mapping method.

\section{Combining the fiber cavity with a high-resolution microscope}

As we described in the previous section, the SNOM tip is a simple and yet powerful tool to realize the mapping of the standing waves inside an open Fabry-Perot cavity. Another key advantage of this method is the fact that after mapping the position of cavity mode, we can use the tip as a nearly perfect point source and align external optical elements with respect to this reference point (Fig.~\ref{fig:Figure4}(a)). It allows also the optimization and characterization of diffraction-limited optical setup.
More specifically, in our experiment, motivated by the prospect of detecting individual cold atoms trapped in the lattice sites by fluorescence, we have combined the previous fiber cavity with a high-NA objective. We have opted for a small, monolithic setup where no adjustment of the objective position will be possible in presence of the atoms. To find the optimal position between the cavity mode and the objective, we first use the SNOM tip to map the cavity mode, whose location cannot be determined with sufficient precision by only centering geometrically with respect to the fiber edges. Then, working as a point source, the tip allows us to assess diffraction-limited operation and to optimize the distance between the objective and the object.

\subsection{Aligning the microscope to the cavity mode}\label{text:tip_microscope}

\begin{figure}[thbp]
    \includegraphics[width=0.49\textwidth]{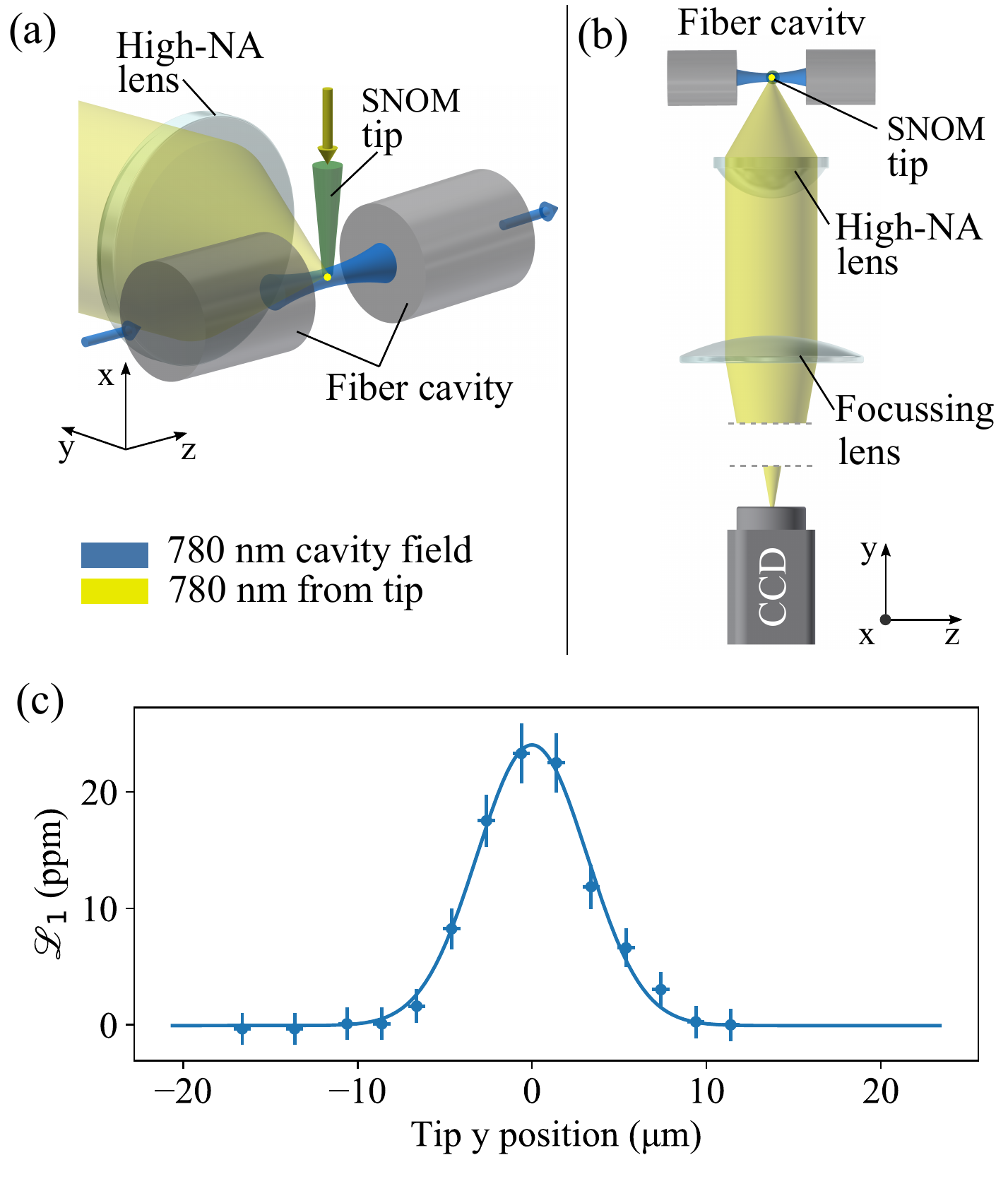}
    \caption{(a) Use of a coated SNOM tip as a point-like source to align a high-NA objective on the optical mode of the fiber Fabry-Perot cavity (not to scale). We inject $780\,$nm light (yellow) in the fiber of the SNOM tip, whose apex aperture constitutes the point-like source. Centering the tip on the $780\,$nm cavity mode (blue) allows aligning the objective with respect to the mode of the resonator. (b) Sketch of the optical setup for imaging the tip light (not to scale). (c) Optical losses at $780\,$nm as a function of the tip displacement along $y$. Dots: experimental data; solid line: best-fit curve using a Gaussian fitting function.}
      \label{fig:Figure4}
\end{figure}

The objective of our microscope is a single commercial molded aspheric lens (Model 352240, LightPath Technologies, Inc.). This lens has a numerical aperture of $0.5$, a focal length of $8\,$mm and a working distance of about $5.7\,$mm, and it has already been employed for single atom trapping and detection.\cite{Sortais2007}
The optical axis of the lens is along the $y$ axis, orthogonal to the $z$ axis of the cavity and to the $x$ axis of the tip (see Fig. \ref{fig:Figure4} (a),(b)). The microscope is then completed with a second lens of $400\,$mm focal length (Thorlabs LA1172), which sets the magnification and allows the image formation on a CCD camera. The high-NA objective is diffraction-limited at $780\,$nm within a range of about $\pm 50\,\upmu$m around the optimal working distance. However, we cannot tolerate variations of the objective-object distance larger than $\pm10\,\upmu$m, as a $10\,\upmu$m variation of the working distance implies a large shift of about $28\,$mm of the image plane, and a consequent change in the magnification.

The SNOM tip is a single-mode tapered optical fiber that, in contrast to the one used previously, is coated with an Aluminium layer (Lovalite, Besançon - FR). An aperture as small as $100\,$nm is present on the tip end and acts as a point-like source for our microscope when light is coupled into the fiber. The increased thickness of the tip due to the coating makes it impossible to resolve the standing-wave modulation of the cavity modes. However, as shown below, it allows probing the transverse section of the modes and to pinpoint their position.
The SNOM tip is fixed on a three-axis translation stage along $x,y,z$ axes and typically few tens of microwatts of $780\,$nm laser light are coupled into the tip fiber.

To align the lens objective with respect to the cavity mode, we implement the following method:
\begin{enumerate}
\item The SNOM tip is inserted in the fiber cavity, until the cavity mode at $780\,$nm is perturbed.
As previously described, we map the transverse profile of the mode by displacing the tip orthogonally to the cavity axis. By monitoring the cavity transmission (Fig.~\ref{fig:Figure4}(c)), we detect the center of the cavity mode as the position of maximal perturbation, with an uncertainty of $\pm2\,\upmu$m due to the limited precision of our manual $y$ translation stage. This sets the $y$ position of the tip.  
\item The position of the tip in the $(x,z)$ plane (parallel to the lens surface) is less critical. We define it to be the geometrical center of the cavity by using an external microscope.
\item Once the position of the tip is set, we use the aperture on the apex as a point-like source for adjusting the objective position, specially the distance between the object and the objective, which has to be adjusted with an accuracy of a few tenth of microns in order to achieve the optimal diffraction-limited operation.    
\end{enumerate}

This method allows to precisely adjust the position of the microscope with respect to the cavity mode in a transitive way, using the SNOM tip as an intermediary reference object.

\subsection{Characterization of the microscope objective with the SNOM tip }

\begin{figure}[thbp]
    \centering
    \includegraphics[width=\columnwidth]{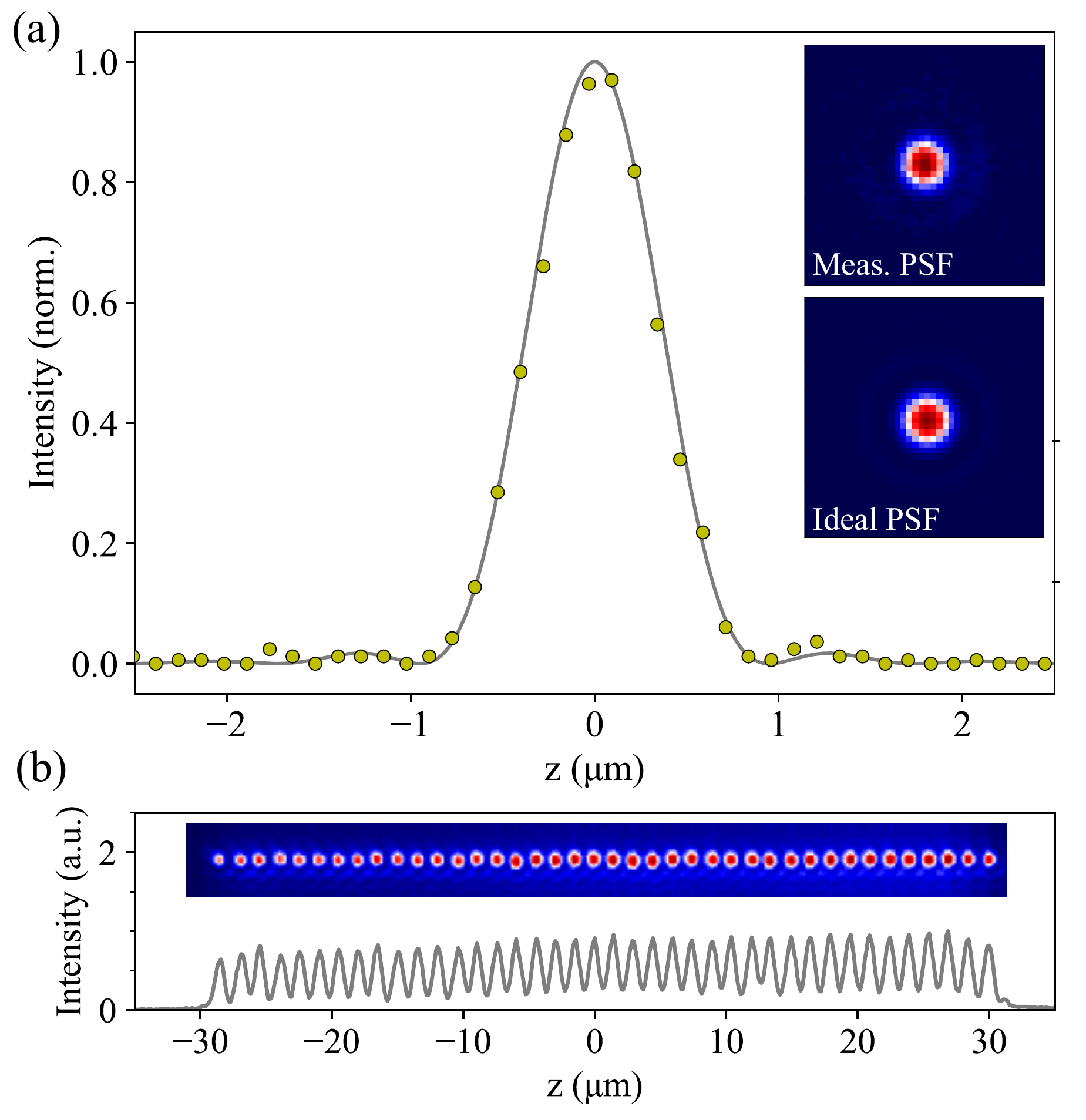}
    \caption{(a) Data points: cut of the PSF measured at the optimal alignment of the high-NA objective. Blue line: ideal Airy profile containing the same optical power. Insets: 2D profiles of the measured and of the ideal PSF. (b) Image and intensity profile of a virtual chain of point-like sources, with $1.5\,\upmu$m spacing. The image is obtained by moving the SNOM tip on the object plane and summing the corresponding PSFs. No re-adjustment of the microscope has been done between the single pictures. The intensity variation of the spot is due to a power fluctuation of the light during the measurement and is not a systematic feature of the imaging system.}
        \label{fig:Figure5}
\end{figure}

In order to check and optimize the performances of the microscope objective described in the previous paragraph, we use the SNOM tip as a reference point-like source. We perform a quantitative study of the point spread function (PSF) of the optical system, defined as the image of a point-like emitter (i.e. with much smaller size than the operating wavelength). In the absence of aberrations, the PSF is the common Airy function. However, in a real optical system, aberrations create deviations of the PSF from the ideal Airy profile. A simple and useful parameter to estimate the amount of aberrations, which only requires a point-like source, is the Strehl ratio defined as $S=I_{\mathrm{aberr}}/I_{\mathrm{ideal}}$
where $I_{\mathrm{aberr}}$ and $I_{\mathrm{ideal}}$ denote the peak intensity of the PSF, respectively with aberrations and in the ideal (stigmatic) case.\cite{Robens2017} An optical system is commonly defined as diffraction-limited when $S\geq0.8$. Fig. \ref{fig:Figure5}(a) shows the PSF at the optimal position, and its comparison with the Airy distribution containing the same optical power: from this comparison, we get a measured value of the Strehl ratio $S>0.95$.  
 
Once the high-NA lens is placed at its optimal position, moving the tip along the optical axis of the cavity allows us to characterize the field of view of the objective. A key requirement for the microscope is indeed the capability of resolving the sites of our intra-cavity lattice along all its extent. We measure that the imaging system shows diffraction-limited operation $S>0.8$ over a range of $\pm25\,\upmu$m, in agreement with Ref.~\onlinecite{Sortais2007}.
For illustration, in Fig.~\ref{fig:Figure5}(b) we show an image obtained by moving the SNOM tip off axis by $\pm30\,\upmu$m with steps of $1.5\,\upmu$m, and adding the individual PSFs. This simulates the operation of the microscope in imaging a 1D array of point-like objects with a spacing that is twice the one of our intra-cavity lattice.

\section{Conclusion}

Our results show that commercially available SNOM tips are an invaluable tool for experiments where the overlap between standing waves has to be determined and/or where a cavity mode has to be aligned to external optical elements such as high-NA objectives. This opens the way to new generation of cavity-QED experiments for quantum technologies where collective coupling of emitters by the cavity mode goes along with individually detection and addressing of the emitter.

\begin{acknowledgments}
We thank Pierre-Antoine Bourdel for careful reading of the manuscript. This project has received funding from:
Agence Nationale de la Recherche (ANR) (SAROCEMA project, ANR-14-CE32-0002);
European Research Council (ERC) under the European Union's Horizon 2020 research and innovation programme Grant agreement No 671133 (EQUEMI project); and the DIM SIRTEQ from Région Ile-de-France.\\
\end{acknowledgments}


\begin{thebibliography}{29}%
\makeatletter
\providecommand \@ifxundefined [1]{%
 \@ifx{#1\undefined}
}%
\providecommand \@ifnum [1]{%
 \ifnum #1\expandafter \@firstoftwo
 \else \expandafter \@secondoftwo
 \fi
}%
\providecommand \@ifx [1]{%
 \ifx #1\expandafter \@firstoftwo
 \else \expandafter \@secondoftwo
 \fi
}%
\providecommand \natexlab [1]{#1}%
\providecommand \enquote  [1]{``#1''}%
\providecommand \bibnamefont  [1]{#1}%
\providecommand \bibfnamefont [1]{#1}%
\providecommand \citenamefont [1]{#1}%
\providecommand \href@noop [0]{\@secondoftwo}%
\providecommand \href [0]{\begingroup \@sanitize@url \@href}%
\providecommand \@href[1]{\@@startlink{#1}\@@href}%
\providecommand \@@href[1]{\endgroup#1\@@endlink}%
\providecommand \@sanitize@url [0]{\catcode `\\12\catcode `\$12\catcode
  `\&12\catcode `\#12\catcode `\^12\catcode `\_12\catcode `\%12\relax}%
\providecommand \@@startlink[1]{}%
\providecommand \@@endlink[0]{}%
\providecommand \url  [0]{\begingroup\@sanitize@url \@url }%
\providecommand \@url [1]{\endgroup\@href {#1}{\urlprefix }}%
\providecommand \urlprefix  [0]{URL }%
\providecommand \Eprint [0]{\href }%
\providecommand \doibase [0]{https://doi.org/}%
\providecommand \selectlanguage [0]{\@gobble}%
\providecommand \bibinfo  [0]{\@secondoftwo}%
\providecommand \bibfield  [0]{\@secondoftwo}%
\providecommand \translation [1]{[#1]}%
\providecommand \BibitemOpen [0]{}%
\providecommand \bibitemStop [0]{}%
\providecommand \bibitemNoStop [0]{.\EOS\space}%
\providecommand \EOS [0]{\spacefactor3000\relax}%
\providecommand \BibitemShut  [1]{\csname bibitem#1\endcsname}%
\let\auto@bib@innerbib\@empty
\bibitem [{\citenamefont {Eisaman}\ \emph {et~al.}(2011)\citenamefont
  {Eisaman}, \citenamefont {Fan}, \citenamefont {Migdall},\ and\ \citenamefont
  {Polyakov}}]{Eisaman2011}%
  \BibitemOpen
  \bibfield  {author} {\bibinfo {author} {\bibfnamefont {M.~D.}\ \bibnamefont
  {Eisaman}}, \bibinfo {author} {\bibfnamefont {J.}~\bibnamefont {Fan}},
  \bibinfo {author} {\bibfnamefont {A.}~\bibnamefont {Migdall}},\ and\ \bibinfo
  {author} {\bibfnamefont {S.~V.}\ \bibnamefont {Polyakov}},\ }\bibfield
  {title} {\enquote {\bibinfo {title} {Invited {{Review Article}}:
  {{Single}}-photon sources and detectors},}\ }\href
  {https://doi.org/10.1063/1.3610677} {\bibfield  {journal} {\bibinfo
  {journal} {Review of Scientific Instruments}\ }\textbf {\bibinfo {volume}
  {82}},\ \bibinfo {pages} {071101} (\bibinfo {year} {2011})}\BibitemShut
  {NoStop}%
\bibitem [{\citenamefont {Heshami}\ \emph {et~al.}(2016)\citenamefont
  {Heshami}, \citenamefont {England}, \citenamefont {Humphreys}, \citenamefont
  {Bustard}, \citenamefont {Acosta}, \citenamefont {Nunn},\ and\ \citenamefont
  {Sussman}}]{Heshami2016}%
  \BibitemOpen
  \bibfield  {author} {\bibinfo {author} {\bibfnamefont {K.}~\bibnamefont
  {Heshami}}, \bibinfo {author} {\bibfnamefont {D.~G.}\ \bibnamefont
  {England}}, \bibinfo {author} {\bibfnamefont {P.~C.}\ \bibnamefont
  {Humphreys}}, \bibinfo {author} {\bibfnamefont {P.~J.}\ \bibnamefont
  {Bustard}}, \bibinfo {author} {\bibfnamefont {V.~M.}\ \bibnamefont {Acosta}},
  \bibinfo {author} {\bibfnamefont {J.}~\bibnamefont {Nunn}},\ and\ \bibinfo
  {author} {\bibfnamefont {B.~J.}\ \bibnamefont {Sussman}},\ }\bibfield
  {title} {\enquote {\bibinfo {title} {Quantum memories: Emerging applications
  and recent advances},}\ }\href
  {https://doi.org/10.1080/09500340.2016.1148212} {\bibfield  {journal}
  {\bibinfo  {journal} {Journal of Modern Optics}\ }\textbf {\bibinfo {volume}
  {63}},\ \bibinfo {pages} {2005--2028} (\bibinfo {year} {2016})}\BibitemShut
  {NoStop}%
\bibitem [{\citenamefont {Benedikter}\ \emph {et~al.}(2015)\citenamefont
  {Benedikter}, \citenamefont {H{\"u}mmer}, \citenamefont {Mader},
  \citenamefont {Schlederer}, \citenamefont {Reichel}, \citenamefont
  {H{\"a}nsch},\ and\ \citenamefont {Hunger}}]{Benedikter2015}%
  \BibitemOpen
  \bibfield  {author} {\bibinfo {author} {\bibfnamefont {J.}~\bibnamefont
  {Benedikter}}, \bibinfo {author} {\bibfnamefont {T.}~\bibnamefont
  {H{\"u}mmer}}, \bibinfo {author} {\bibfnamefont {M.}~\bibnamefont {Mader}},
  \bibinfo {author} {\bibfnamefont {B.}~\bibnamefont {Schlederer}}, \bibinfo
  {author} {\bibfnamefont {J.}~\bibnamefont {Reichel}}, \bibinfo {author}
  {\bibfnamefont {T.~W.}\ \bibnamefont {H{\"a}nsch}},\ and\ \bibinfo {author}
  {\bibfnamefont {D.}~\bibnamefont {Hunger}},\ }\bibfield  {title} {\enquote
  {\bibinfo {title} {Transverse-mode coupling and diffraction loss in tunable
  {{Fabry}}\textendash{{P{\'e}rot}} microcavities},}\ }\href
  {https://doi.org/10.1088/1367-2630/17/5/053051} {\bibfield  {journal}
  {\bibinfo  {journal} {New Journal of Physics}\ }\textbf {\bibinfo {volume}
  {17}},\ \bibinfo {pages} {053051} (\bibinfo {year} {2015})}\BibitemShut
  {NoStop}%
\bibitem [{\citenamefont {Hosten}\ \emph {et~al.}(2016)\citenamefont {Hosten},
  \citenamefont {Engelsen}, \citenamefont {Krishnakumar},\ and\ \citenamefont
  {Kasevich}}]{Hosten2016}%
  \BibitemOpen
  \bibfield  {author} {\bibinfo {author} {\bibfnamefont {O.}~\bibnamefont
  {Hosten}}, \bibinfo {author} {\bibfnamefont {N.~J.}\ \bibnamefont
  {Engelsen}}, \bibinfo {author} {\bibfnamefont {R.}~\bibnamefont
  {Krishnakumar}},\ and\ \bibinfo {author} {\bibfnamefont {M.~A.}\ \bibnamefont
  {Kasevich}},\ }\bibfield  {title} {\enquote {\bibinfo {title} {Measurement
  noise 100 times lower than the quantum-projection limit using entangled
  atoms},}\ }\href {https://doi.org/10.1038/nature16176} {\bibfield  {journal}
  {\bibinfo  {journal} {Nature}\ }\textbf {\bibinfo {volume} {529}},\ \bibinfo
  {pages} {505--508} (\bibinfo {year} {2016})}\BibitemShut {NoStop}%
\bibitem [{\citenamefont {Zhiqiang}\ \emph {et~al.}(2017)\citenamefont
  {Zhiqiang}, \citenamefont {Lee}, \citenamefont {Kumar}, \citenamefont
  {Arnold}, \citenamefont {Masson}, \citenamefont {Parkins},\ and\
  \citenamefont {Barrett}}]{Zhiqiang2017}%
  \BibitemOpen
  \bibfield  {author} {\bibinfo {author} {\bibfnamefont {Z.}~\bibnamefont
  {Zhiqiang}}, \bibinfo {author} {\bibfnamefont {C.~H.}\ \bibnamefont {Lee}},
  \bibinfo {author} {\bibfnamefont {R.}~\bibnamefont {Kumar}}, \bibinfo
  {author} {\bibfnamefont {K.~J.}\ \bibnamefont {Arnold}}, \bibinfo {author}
  {\bibfnamefont {S.~J.}\ \bibnamefont {Masson}}, \bibinfo {author}
  {\bibfnamefont {A.~S.}\ \bibnamefont {Parkins}},\ and\ \bibinfo {author}
  {\bibfnamefont {M.~D.}\ \bibnamefont {Barrett}},\ }\bibfield  {title}
  {\enquote {\bibinfo {title} {Nonequilibrium phase transition in a spin-1
  {{Dicke}} model},}\ }\href {https://doi.org/10.1364/OPTICA.4.000424}
  {\bibfield  {journal} {\bibinfo  {journal} {Optica}\ }\textbf {\bibinfo
  {volume} {4}},\ \bibinfo {pages} {424--429} (\bibinfo {year}
  {2017})}\BibitemShut {NoStop}%
\bibitem [{\citenamefont {Davis}\ \emph {et~al.}(2019)\citenamefont {Davis},
  \citenamefont {Bentsen}, \citenamefont {Homeier}, \citenamefont {Li},\ and\
  \citenamefont {{Schleier-Smith}}}]{Davis2019}%
  \BibitemOpen
  \bibfield  {author} {\bibinfo {author} {\bibfnamefont {E.~J.}\ \bibnamefont
  {Davis}}, \bibinfo {author} {\bibfnamefont {G.}~\bibnamefont {Bentsen}},
  \bibinfo {author} {\bibfnamefont {L.}~\bibnamefont {Homeier}}, \bibinfo
  {author} {\bibfnamefont {T.}~\bibnamefont {Li}},\ and\ \bibinfo {author}
  {\bibfnamefont {M.~H.}\ \bibnamefont {{Schleier-Smith}}},\ }\bibfield
  {title} {\enquote {\bibinfo {title} {Photon-{{Mediated Spin}}-{{Exchange
  Dynamics}} of {{Spin}}-1 {{Atoms}}},}\ }\href
  {https://doi.org/10.1103/PhysRevLett.122.010405} {\bibfield  {journal}
  {\bibinfo  {journal} {Physical Review Letters}\ }\textbf {\bibinfo {volume}
  {122}},\ \bibinfo {pages} {010405} (\bibinfo {year} {2019})}\BibitemShut
  {NoStop}%
\bibitem [{\citenamefont {Garcia}\ \emph {et~al.}(2018)\citenamefont {Garcia},
  \citenamefont {Ferri}, \citenamefont {Ott}, \citenamefont {Reichel},\ and\
  \citenamefont {Long}}]{Garcia2018}%
  \BibitemOpen
  \bibfield  {author} {\bibinfo {author} {\bibfnamefont {S.}~\bibnamefont
  {Garcia}}, \bibinfo {author} {\bibfnamefont {F.}~\bibnamefont {Ferri}},
  \bibinfo {author} {\bibfnamefont {K.}~\bibnamefont {Ott}}, \bibinfo {author}
  {\bibfnamefont {J.}~\bibnamefont {Reichel}},\ and\ \bibinfo {author}
  {\bibfnamefont {R.}~\bibnamefont {Long}},\ }\bibfield  {title} {\enquote
  {\bibinfo {title} {Dual-wavelength fiber {{Fabry}}-{{Perot}} cavities with
  engineered birefringence},}\ }\href {https://doi.org/10.1364/OE.26.022249}
  {\bibfield  {journal} {\bibinfo  {journal} {Optics Express}\ }\textbf
  {\bibinfo {volume} {26}},\ \bibinfo {pages} {22249} (\bibinfo {year}
  {2018})}\BibitemShut {NoStop}%
\bibitem [{\citenamefont {Welte}\ \emph {et~al.}(2017)\citenamefont {Welte},
  \citenamefont {Hacker}, \citenamefont {Daiss}, \citenamefont {Ritter},\ and\
  \citenamefont {Rempe}}]{Welte2017}%
  \BibitemOpen
  \bibfield  {author} {\bibinfo {author} {\bibfnamefont {S.}~\bibnamefont
  {Welte}}, \bibinfo {author} {\bibfnamefont {B.}~\bibnamefont {Hacker}},
  \bibinfo {author} {\bibfnamefont {S.}~\bibnamefont {Daiss}}, \bibinfo
  {author} {\bibfnamefont {S.}~\bibnamefont {Ritter}},\ and\ \bibinfo {author}
  {\bibfnamefont {G.}~\bibnamefont {Rempe}},\ }\bibfield  {title} {\enquote
  {\bibinfo {title} {Cavity {{Carving}} of {{Atomic Bell States}}},}\ }\href
  {https://doi.org/10.1103/PhysRevLett.118.210503} {\bibfield  {journal}
  {\bibinfo  {journal} {Physical Review Letters}\ }\textbf {\bibinfo {volume}
  {118}},\ \bibinfo {pages} {210503} (\bibinfo {year} {2017})}\BibitemShut
  {NoStop}%
\bibitem [{\citenamefont {Léonard}(2017)}]{Leonard2017}%
  \BibitemOpen
  \bibfield  {author} {\bibinfo {author} {\bibfnamefont {J.}~\bibnamefont
  {Léonard}},\ }\emph {\bibinfo {title} {A Supersolid of Matter and Light}},\
  \href@noop {} {Ph.D. thesis},\ \bibinfo  {school} {ETH Z\"{u}rich} (\bibinfo
  {year} {2017})\BibitemShut {NoStop}%
\bibitem [{\citenamefont {Gallego}\ \emph {et~al.}(2018)\citenamefont
  {Gallego}, \citenamefont {Alt}, \citenamefont {Macha}, \citenamefont
  {{Martinez-Dorantes}}, \citenamefont {Pandey},\ and\ \citenamefont
  {Meschede}}]{Gallego2018}%
  \BibitemOpen
  \bibfield  {author} {\bibinfo {author} {\bibfnamefont {J.}~\bibnamefont
  {Gallego}}, \bibinfo {author} {\bibfnamefont {W.}~\bibnamefont {Alt}},
  \bibinfo {author} {\bibfnamefont {T.}~\bibnamefont {Macha}}, \bibinfo
  {author} {\bibfnamefont {M.}~\bibnamefont {{Martinez-Dorantes}}}, \bibinfo
  {author} {\bibfnamefont {D.}~\bibnamefont {Pandey}},\ and\ \bibinfo {author}
  {\bibfnamefont {D.}~\bibnamefont {Meschede}},\ }\bibfield  {title} {\enquote
  {\bibinfo {title} {Strong {{Purcell Effect}} on a {{Neutral Atom Trapped}} in
  an {{Open Fiber Cavity}}},}\ }\href
  {https://doi.org/10.1103/PhysRevLett.121.173603} {\bibfield  {journal}
  {\bibinfo  {journal} {Physical Review Letters}\ }\textbf {\bibinfo {volume}
  {121}},\ \bibinfo {pages} {173603} (\bibinfo {year} {2018})}\BibitemShut
  {NoStop}%
\bibitem [{\citenamefont {Wiener}(1890)}]{Wiener1890}%
  \BibitemOpen
  \bibfield  {author} {\bibinfo {author} {\bibfnamefont {O.}~\bibnamefont
  {Wiener}},\ }\bibfield  {title} {\enquote {\bibinfo {title} {Stehende
  {{Lichtwellen}} und die {{Schwingungsrichtung}} polarisirten {{Lichtes}}},}\
  }\href {https://doi.org/10.1002/andp.18902760603} {\bibfield  {journal}
  {\bibinfo  {journal} {Annalen der Physik}\ }\textbf {\bibinfo {volume}
  {276}},\ \bibinfo {pages} {203--243} (\bibinfo {year} {1890})}\BibitemShut
  {NoStop}%
\bibitem [{\citenamefont {Sasaki}, \citenamefont {Mi},\ and\ \citenamefont
  {Hane}(1999)}]{Sasaki1999}%
  \BibitemOpen
  \bibfield  {author} {\bibinfo {author} {\bibfnamefont {M.}~\bibnamefont
  {Sasaki}}, \bibinfo {author} {\bibfnamefont {X.}~\bibnamefont {Mi}},\ and\
  \bibinfo {author} {\bibfnamefont {K.}~\bibnamefont {Hane}},\ }\bibfield
  {title} {\enquote {\bibinfo {title} {Standing wave detection and
  interferometer application using a photodiode thinner than optical
  wavelength},}\ }\href {https://doi.org/10.1063/1.124898} {\bibfield
  {journal} {\bibinfo  {journal} {Applied Physics Letters}\ }\textbf {\bibinfo
  {volume} {75}},\ \bibinfo {pages} {2008--2010} (\bibinfo {year}
  {1999})}\BibitemShut {NoStop}%
\bibitem [{\citenamefont {Stiebig}, \citenamefont {Knipp},\ and\ \citenamefont
  {Bunte}(2006)}]{Stiebig2006}%
  \BibitemOpen
  \bibfield  {author} {\bibinfo {author} {\bibfnamefont {H.}~\bibnamefont
  {Stiebig}}, \bibinfo {author} {\bibfnamefont {D.}~\bibnamefont {Knipp}},\
  and\ \bibinfo {author} {\bibfnamefont {E.}~\bibnamefont {Bunte}},\ }\bibfield
   {title} {\enquote {\bibinfo {title} {Standing-wave spectrometer},}\ }\href
  {https://doi.org/10.1063/1.2179610} {\bibfield  {journal} {\bibinfo
  {journal} {Applied Physics Letters}\ }\textbf {\bibinfo {volume} {88}},\
  \bibinfo {pages} {083509} (\bibinfo {year} {2006})}\BibitemShut {NoStop}%
\bibitem [{\citenamefont {Knight}\ \emph {et~al.}(1996)\citenamefont {Knight},
  \citenamefont {Dubreuil}, \citenamefont {Sandoghdar}, \citenamefont {Hare},
  \citenamefont {{Lef{\`e}vre-Seguin}}, \citenamefont {Raimond},\ and\
  \citenamefont {Haroche}}]{Knight1996}%
  \BibitemOpen
  \bibfield  {author} {\bibinfo {author} {\bibfnamefont {J.~C.}\ \bibnamefont
  {Knight}}, \bibinfo {author} {\bibfnamefont {N.}~\bibnamefont {Dubreuil}},
  \bibinfo {author} {\bibfnamefont {V.}~\bibnamefont {Sandoghdar}}, \bibinfo
  {author} {\bibfnamefont {J.}~\bibnamefont {Hare}}, \bibinfo {author}
  {\bibfnamefont {V.}~\bibnamefont {{Lef{\`e}vre-Seguin}}}, \bibinfo {author}
  {\bibfnamefont {J.~M.}\ \bibnamefont {Raimond}},\ and\ \bibinfo {author}
  {\bibfnamefont {S.}~\bibnamefont {Haroche}},\ }\bibfield  {title} {\enquote
  {\bibinfo {title} {Characterizing whispering-gallery modes in microspheres by
  direct observation of the optical standing-wave pattern in the near field},}\
  }\href {https://doi.org/10.1364/OL.21.000698} {\bibfield  {journal} {\bibinfo
   {journal} {Optics Letters}\ }\textbf {\bibinfo {volume} {21}},\ \bibinfo
  {pages} {698} (\bibinfo {year} {1996})}\BibitemShut {NoStop}%
\bibitem [{\citenamefont {Gotzinger}\ \emph {et~al.}(2001)\citenamefont
  {Gotzinger}, \citenamefont {Demmerer}, \citenamefont {Benson},\ and\
  \citenamefont {Sandoghdar}}]{Gotzinger2001}%
  \BibitemOpen
  \bibfield  {author} {\bibinfo {author} {\bibfnamefont {S.}~\bibnamefont
  {Gotzinger}}, \bibinfo {author} {\bibfnamefont {S.}~\bibnamefont {Demmerer}},
  \bibinfo {author} {\bibfnamefont {O.}~\bibnamefont {Benson}},\ and\ \bibinfo
  {author} {\bibfnamefont {V.}~\bibnamefont {Sandoghdar}},\ }\bibfield  {title}
  {\enquote {\bibinfo {title} {Mapping and manipulating whispering gallery
  modes of a microsphere resonator with a near-field probe},}\ }\href
  {https://doi.org/10.1046/j.1365-2818.2001.00865.x} {\bibfield  {journal}
  {\bibinfo  {journal} {Journal of Microscopy}\ }\textbf {\bibinfo {volume}
  {202}},\ \bibinfo {pages} {117--121} (\bibinfo {year} {2001})}\BibitemShut
  {NoStop}%
\bibitem [{\citenamefont {Koenderink}\ \emph {et~al.}(2005)\citenamefont
  {Koenderink}, \citenamefont {Kafesaki}, \citenamefont {Buchler},\ and\
  \citenamefont {Sandoghdar}}]{Koenderink2005}%
  \BibitemOpen
  \bibfield  {author} {\bibinfo {author} {\bibfnamefont {A.~F.}\ \bibnamefont
  {Koenderink}}, \bibinfo {author} {\bibfnamefont {M.}~\bibnamefont
  {Kafesaki}}, \bibinfo {author} {\bibfnamefont {B.~C.}\ \bibnamefont
  {Buchler}},\ and\ \bibinfo {author} {\bibfnamefont {V.}~\bibnamefont
  {Sandoghdar}},\ }\bibfield  {title} {\enquote {\bibinfo {title} {Controlling
  the {{Resonance}} of a {{Photonic Crystal Microcavity}} by a {{Near}}-{{Field
  Probe}}},}\ }\href {https://doi.org/10.1103/PhysRevLett.95.153904} {\bibfield
   {journal} {\bibinfo  {journal} {Physical Review Letters}\ }\textbf {\bibinfo
  {volume} {95}},\ \bibinfo {pages} {153904} (\bibinfo {year}
  {2005})}\BibitemShut {NoStop}%
\bibitem [{\citenamefont {Rotenberg}\ and\ \citenamefont
  {Kuipers}(2014)}]{Rotenberg2014}%
  \BibitemOpen
  \bibfield  {author} {\bibinfo {author} {\bibfnamefont {N.}~\bibnamefont
  {Rotenberg}}\ and\ \bibinfo {author} {\bibfnamefont {L.}~\bibnamefont
  {Kuipers}},\ }\bibfield  {title} {\enquote {\bibinfo {title} {Mapping
  nanoscale light fields},}\ }\href {https://doi.org/10.1038/nphoton.2014.285}
  {\bibfield  {journal} {\bibinfo  {journal} {Nature Photonics}\ }\textbf
  {\bibinfo {volume} {8}},\ \bibinfo {pages} {919--926} (\bibinfo {year}
  {2014})}\BibitemShut {NoStop}%
\bibitem [{\citenamefont {Courjon}, \citenamefont {Bainier},\ and\
  \citenamefont {Baida}(1994)}]{Courjon1994}%
  \BibitemOpen
  \bibfield  {author} {\bibinfo {author} {\bibfnamefont {D.}~\bibnamefont
  {Courjon}}, \bibinfo {author} {\bibfnamefont {C.}~\bibnamefont {Bainier}},\
  and\ \bibinfo {author} {\bibfnamefont {F.}~\bibnamefont {Baida}},\ }\bibfield
   {title} {\enquote {\bibinfo {title} {Seeing inside a {{Fabry}}-{{P{\'e}rot}}
  resonator by means of a scanning tunneling optical microscope},}\ }\href
  {https://doi.org/10.1016/0030-4018(94)90162-7} {\bibfield  {journal}
  {\bibinfo  {journal} {Optics Communications}\ }\textbf {\bibinfo {volume}
  {110}},\ \bibinfo {pages} {7--12} (\bibinfo {year} {1994})}\BibitemShut
  {NoStop}%
\bibitem [{\citenamefont {Guth{\"o}hrlein}\ \emph {et~al.}(2001)\citenamefont
  {Guth{\"o}hrlein}, \citenamefont {Keller}, \citenamefont {Hayasaka},
  \citenamefont {Lange},\ and\ \citenamefont {Walther}}]{Guthohrlein2001}%
  \BibitemOpen
  \bibfield  {author} {\bibinfo {author} {\bibfnamefont {G.~R.}\ \bibnamefont
  {Guth{\"o}hrlein}}, \bibinfo {author} {\bibfnamefont {M.}~\bibnamefont
  {Keller}}, \bibinfo {author} {\bibfnamefont {K.}~\bibnamefont {Hayasaka}},
  \bibinfo {author} {\bibfnamefont {W.}~\bibnamefont {Lange}},\ and\ \bibinfo
  {author} {\bibfnamefont {H.}~\bibnamefont {Walther}},\ }\bibfield  {title}
  {\enquote {\bibinfo {title} {A single ion as a nanoscopic probe of an optical
  field},}\ }\href {https://doi.org/10.1038/35102129} {\bibfield  {journal}
  {\bibinfo  {journal} {Nature}\ }\textbf {\bibinfo {volume} {414}},\ \bibinfo
  {pages} {49--51} (\bibinfo {year} {2001})}\BibitemShut {NoStop}%
\bibitem [{\citenamefont {Favero}\ \emph {et~al.}(2009)\citenamefont {Favero},
  \citenamefont {Stapfner}, \citenamefont {Hunger}, \citenamefont
  {Paulitschke}, \citenamefont {Reichel}, \citenamefont {Lorenz}, \citenamefont
  {Weig},\ and\ \citenamefont {Karrai}}]{Favero2009}%
  \BibitemOpen
  \bibfield  {author} {\bibinfo {author} {\bibfnamefont {I.}~\bibnamefont
  {Favero}}, \bibinfo {author} {\bibfnamefont {S.}~\bibnamefont {Stapfner}},
  \bibinfo {author} {\bibfnamefont {D.}~\bibnamefont {Hunger}}, \bibinfo
  {author} {\bibfnamefont {P.}~\bibnamefont {Paulitschke}}, \bibinfo {author}
  {\bibfnamefont {J.}~\bibnamefont {Reichel}}, \bibinfo {author} {\bibfnamefont
  {H.}~\bibnamefont {Lorenz}}, \bibinfo {author} {\bibfnamefont {E.~M.}\
  \bibnamefont {Weig}},\ and\ \bibinfo {author} {\bibfnamefont
  {K.}~\bibnamefont {Karrai}},\ }\bibfield  {title} {\enquote {\bibinfo {title}
  {Fluctuating nanomechanical system in a high finesse optical microcavity},}\
  }\href {https://doi.org/10.1364/OE.17.012813} {\bibfield  {journal} {\bibinfo
   {journal} {Optics Express}\ }\textbf {\bibinfo {volume} {17}},\ \bibinfo
  {pages} {12813} (\bibinfo {year} {2009})}\BibitemShut {NoStop}%
\bibitem [{\citenamefont {Gloppe}\ \emph {et~al.}(2014)\citenamefont {Gloppe},
  \citenamefont {Verlot}, \citenamefont {{Dupont-Ferrier}}, \citenamefont
  {Siria}, \citenamefont {Poncharal}, \citenamefont {Bachelier}, \citenamefont
  {Vincent},\ and\ \citenamefont {Arcizet}}]{Gloppe2014}%
  \BibitemOpen
  \bibfield  {author} {\bibinfo {author} {\bibfnamefont {A.}~\bibnamefont
  {Gloppe}}, \bibinfo {author} {\bibfnamefont {P.}~\bibnamefont {Verlot}},
  \bibinfo {author} {\bibfnamefont {E.}~\bibnamefont {{Dupont-Ferrier}}},
  \bibinfo {author} {\bibfnamefont {A.}~\bibnamefont {Siria}}, \bibinfo
  {author} {\bibfnamefont {P.}~\bibnamefont {Poncharal}}, \bibinfo {author}
  {\bibfnamefont {G.}~\bibnamefont {Bachelier}}, \bibinfo {author}
  {\bibfnamefont {P.}~\bibnamefont {Vincent}},\ and\ \bibinfo {author}
  {\bibfnamefont {O.}~\bibnamefont {Arcizet}},\ }\bibfield  {title} {\enquote
  {\bibinfo {title} {Bidimensional nano-optomechanics and topological
  backaction in a non-conservative radiation force field},}\ }\href
  {https://doi.org/10.1038/nnano.2014.189} {\bibfield  {journal} {\bibinfo
  {journal} {Nature Nanotechnology}\ }\textbf {\bibinfo {volume} {9}},\
  \bibinfo {pages} {920--926} (\bibinfo {year} {2014})}\BibitemShut {NoStop}%
\bibitem [{\citenamefont {Fogliano}\ \emph {et~al.}(2019)\citenamefont
  {Fogliano}, \citenamefont {Besga}, \citenamefont {Reigue}, \citenamefont
  {Heringlake}, \citenamefont {{de L{\'e}pinay}}, \citenamefont {Vaneph},
  \citenamefont {Reichel}, \citenamefont {Pigeau},\ and\ \citenamefont
  {Arcizet}}]{Fogliano2019}%
  \BibitemOpen
  \bibfield  {author} {\bibinfo {author} {\bibfnamefont {F.}~\bibnamefont
  {Fogliano}}, \bibinfo {author} {\bibfnamefont {B.}~\bibnamefont {Besga}},
  \bibinfo {author} {\bibfnamefont {A.}~\bibnamefont {Reigue}}, \bibinfo
  {author} {\bibfnamefont {P.}~\bibnamefont {Heringlake}}, \bibinfo {author}
  {\bibfnamefont {L.~M.}\ \bibnamefont {{de L{\'e}pinay}}}, \bibinfo {author}
  {\bibfnamefont {C.}~\bibnamefont {Vaneph}}, \bibinfo {author} {\bibfnamefont
  {J.}~\bibnamefont {Reichel}}, \bibinfo {author} {\bibfnamefont
  {B.}~\bibnamefont {Pigeau}},\ and\ \bibinfo {author} {\bibfnamefont
  {O.}~\bibnamefont {Arcizet}},\ }\bibfield  {title} {\enquote {\bibinfo
  {title} {Cavity nano-optomechanics in the ultrastrong coupling regime with
  ultrasensitive force sensors},}\ }\href {http://arxiv.org/abs/1904.01140} {\
  (\bibinfo {year} {2019})},\ \Eprint {https://arxiv.org/abs/1904.01140}
  {arXiv:1904.01140 [quant-ph]} \BibitemShut {NoStop}%
\bibitem [{\citenamefont {Thompson}\ \emph {et~al.}(2008)\citenamefont
  {Thompson}, \citenamefont {Zwickl}, \citenamefont {Jayich}, \citenamefont
  {Marquardt}, \citenamefont {Girvin},\ and\ \citenamefont
  {Harris}}]{Thompson2008}%
  \BibitemOpen
  \bibfield  {author} {\bibinfo {author} {\bibfnamefont {J.~D.}\ \bibnamefont
  {Thompson}}, \bibinfo {author} {\bibfnamefont {B.~M.}\ \bibnamefont
  {Zwickl}}, \bibinfo {author} {\bibfnamefont {A.~M.}\ \bibnamefont {Jayich}},
  \bibinfo {author} {\bibfnamefont {F.}~\bibnamefont {Marquardt}}, \bibinfo
  {author} {\bibfnamefont {S.~M.}\ \bibnamefont {Girvin}},\ and\ \bibinfo
  {author} {\bibfnamefont {J.~G.~E.}\ \bibnamefont {Harris}},\ }\bibfield
  {title} {\enquote {\bibinfo {title} {Strong dispersive coupling of a
  high-finesse cavity to a micromechanical membrane},}\ }\href
  {https://doi.org/10.1038/nature06715} {\bibfield  {journal} {\bibinfo
  {journal} {Nature}\ }\textbf {\bibinfo {volume} {452}},\ \bibinfo {pages}
  {72--75} (\bibinfo {year} {2008})}\BibitemShut {NoStop}%
\bibitem [{\citenamefont {Vochezer}\ \emph {et~al.}(2018)\citenamefont
  {Vochezer}, \citenamefont {Kampschulte}, \citenamefont {Hammerer},\ and\
  \citenamefont {Treutlein}}]{Vochezer2018}%
  \BibitemOpen
  \bibfield  {author} {\bibinfo {author} {\bibfnamefont {A.}~\bibnamefont
  {Vochezer}}, \bibinfo {author} {\bibfnamefont {T.}~\bibnamefont
  {Kampschulte}}, \bibinfo {author} {\bibfnamefont {K.}~\bibnamefont
  {Hammerer}},\ and\ \bibinfo {author} {\bibfnamefont {P.}~\bibnamefont
  {Treutlein}},\ }\bibfield  {title} {\enquote {\bibinfo {title}
  {Light-{{Mediated Collective Atomic Motion}} in an {{Optical Lattice
  Coupled}} to a {{Membrane}}},}\ }\href
  {https://doi.org/10.1103/PhysRevLett.120.073602} {\bibfield  {journal}
  {\bibinfo  {journal} {Physical Review Letters}\ }\textbf {\bibinfo {volume}
  {120}},\ \bibinfo {pages} {73602} (\bibinfo {year} {2018})}\BibitemShut
  {NoStop}%
\bibitem [{\citenamefont {Hunger}\ \emph {et~al.}(2010)\citenamefont {Hunger},
  \citenamefont {Steinmetz}, \citenamefont {Colombe}, \citenamefont {Deutsch},
  \citenamefont {H{\"a}nsch},\ and\ \citenamefont {Reichel}}]{Hunger2010}%
  \BibitemOpen
  \bibfield  {author} {\bibinfo {author} {\bibfnamefont {D.}~\bibnamefont
  {Hunger}}, \bibinfo {author} {\bibfnamefont {T.}~\bibnamefont {Steinmetz}},
  \bibinfo {author} {\bibfnamefont {Y.}~\bibnamefont {Colombe}}, \bibinfo
  {author} {\bibfnamefont {C.}~\bibnamefont {Deutsch}}, \bibinfo {author}
  {\bibfnamefont {T.~W.}\ \bibnamefont {H{\"a}nsch}},\ and\ \bibinfo {author}
  {\bibfnamefont {J.}~\bibnamefont {Reichel}},\ }\bibfield  {title} {\enquote
  {\bibinfo {title} {A fiber {{Fabry}}\textendash{{Perot}} cavity with high
  finesse},}\ }\href {https://doi.org/10.1088/1367-2630/12/6/065038} {\bibfield
   {journal} {\bibinfo  {journal} {New Journal of Physics}\ }\textbf {\bibinfo
  {volume} {12}},\ \bibinfo {pages} {065038} (\bibinfo {year}
  {2010})}\BibitemShut {NoStop}%
\bibitem [{\citenamefont {Barontini}\ \emph {et~al.}(2015)\citenamefont
  {Barontini}, \citenamefont {Hohmann}, \citenamefont {Haas}, \citenamefont
  {Esteve},\ and\ \citenamefont {Reichel}}]{Barontini2015}%
  \BibitemOpen
  \bibfield  {author} {\bibinfo {author} {\bibfnamefont {G.}~\bibnamefont
  {Barontini}}, \bibinfo {author} {\bibfnamefont {L.}~\bibnamefont {Hohmann}},
  \bibinfo {author} {\bibfnamefont {F.}~\bibnamefont {Haas}}, \bibinfo {author}
  {\bibfnamefont {J.}~\bibnamefont {Esteve}},\ and\ \bibinfo {author}
  {\bibfnamefont {J.}~\bibnamefont {Reichel}},\ }\bibfield  {title} {\enquote
  {\bibinfo {title} {Deterministic generation of multiparticle entanglement by
  quantum {{Zeno}} dynamics},}\ }\href
  {https://doi.org/10.1126/science.aaa0754} {\bibfield  {journal} {\bibinfo
  {journal} {Science}\ }\textbf {\bibinfo {volume} {349}},\ \bibinfo {pages}
  {1317--1321} (\bibinfo {year} {2015})}\BibitemShut {NoStop}%
\bibitem [{\citenamefont {McConnell}\ \emph {et~al.}(2015)\citenamefont
  {McConnell}, \citenamefont {Zhang}, \citenamefont {Hu}, \citenamefont
  {{\'C}uk},\ and\ \citenamefont {Vuleti{\'c}}}]{McConnell2015}%
  \BibitemOpen
  \bibfield  {author} {\bibinfo {author} {\bibfnamefont {R.}~\bibnamefont
  {McConnell}}, \bibinfo {author} {\bibfnamefont {H.}~\bibnamefont {Zhang}},
  \bibinfo {author} {\bibfnamefont {J.}~\bibnamefont {Hu}}, \bibinfo {author}
  {\bibfnamefont {S.}~\bibnamefont {{\'C}uk}},\ and\ \bibinfo {author}
  {\bibfnamefont {V.}~\bibnamefont {Vuleti{\'c}}},\ }\bibfield  {title}
  {\enquote {\bibinfo {title} {Entanglement with negative {{Wigner}} function
  of almost 3,000 atoms heralded by one photon},}\ }\href
  {https://doi.org/10.1038/nature14293} {\bibfield  {journal} {\bibinfo
  {journal} {Nature}\ }\textbf {\bibinfo {volume} {519}},\ \bibinfo {pages}
  {439--442} (\bibinfo {year} {2015})}\BibitemShut {NoStop}%
\bibitem [{\citenamefont {Sortais}\ \emph {et~al.}(2007)\citenamefont
  {Sortais}, \citenamefont {Marion}, \citenamefont {Tuchendler}, \citenamefont
  {Lance}, \citenamefont {Lamare}, \citenamefont {Fournet}, \citenamefont
  {Armellin}, \citenamefont {Mercier}, \citenamefont {Messin}, \citenamefont
  {Browaeys},\ and\ \citenamefont {Grangier}}]{Sortais2007}%
  \BibitemOpen
  \bibfield  {author} {\bibinfo {author} {\bibfnamefont {Y.~R.~P.}\
  \bibnamefont {Sortais}}, \bibinfo {author} {\bibfnamefont {H.}~\bibnamefont
  {Marion}}, \bibinfo {author} {\bibfnamefont {C.}~\bibnamefont {Tuchendler}},
  \bibinfo {author} {\bibfnamefont {A.~M.}\ \bibnamefont {Lance}}, \bibinfo
  {author} {\bibfnamefont {M.}~\bibnamefont {Lamare}}, \bibinfo {author}
  {\bibfnamefont {P.}~\bibnamefont {Fournet}}, \bibinfo {author} {\bibfnamefont
  {C.}~\bibnamefont {Armellin}}, \bibinfo {author} {\bibfnamefont
  {R.}~\bibnamefont {Mercier}}, \bibinfo {author} {\bibfnamefont
  {G.}~\bibnamefont {Messin}}, \bibinfo {author} {\bibfnamefont
  {A.}~\bibnamefont {Browaeys}},\ and\ \bibinfo {author} {\bibfnamefont
  {P.}~\bibnamefont {Grangier}},\ }\bibfield  {title} {\enquote {\bibinfo
  {title} {Diffraction-limited optics for single-atom manipulation},}\ }\href
  {https://doi.org/10.1103/PhysRevA.75.013406} {\bibfield  {journal} {\bibinfo
  {journal} {Physical Review A}\ }\textbf {\bibinfo {volume} {75}},\ \bibinfo
  {pages} {013406} (\bibinfo {year} {2007})}\BibitemShut {NoStop}%
\bibitem [{\citenamefont {Robens}\ \emph {et~al.}(2017)\citenamefont {Robens},
  \citenamefont {Brakhane}, \citenamefont {Alt}, \citenamefont {Klei{\ss}ler},
  \citenamefont {Meschede}, \citenamefont {Moon}, \citenamefont {Ramola},\ and\
  \citenamefont {Alberti}}]{Robens2017}%
  \BibitemOpen
  \bibfield  {author} {\bibinfo {author} {\bibfnamefont {C.}~\bibnamefont
  {Robens}}, \bibinfo {author} {\bibfnamefont {S.}~\bibnamefont {Brakhane}},
  \bibinfo {author} {\bibfnamefont {W.}~\bibnamefont {Alt}}, \bibinfo {author}
  {\bibfnamefont {F.}~\bibnamefont {Klei{\ss}ler}}, \bibinfo {author}
  {\bibfnamefont {D.}~\bibnamefont {Meschede}}, \bibinfo {author}
  {\bibfnamefont {G.}~\bibnamefont {Moon}}, \bibinfo {author} {\bibfnamefont
  {G.}~\bibnamefont {Ramola}},\ and\ \bibinfo {author} {\bibfnamefont
  {A.}~\bibnamefont {Alberti}},\ }\bibfield  {title} {\enquote {\bibinfo
  {title} {A high numerical aperture ({{NA}} = 0.92) objective lens for imaging
  and addressing of cold atoms},}\ }\href
  {https://doi.org/10.1364/OL.42.001043} {\bibfield  {journal} {\bibinfo
  {journal} {Optics Letters}\ }\textbf {\bibinfo {volume} {42}},\ \bibinfo
  {pages} {1043} (\bibinfo {year} {2017})}\BibitemShut {NoStop}%
\end{thebibliography}
%

\end{document}